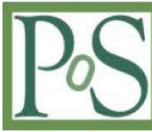

# US Accelerator R&D Program Toward Intensity Frontier Machines

**Vladimir Shiltsev**[1]

*Fermi National Accelerator Laboratory*
*PO Box 500, Batavia IL 60510, USA*
*E-mail:* `shiltsev@fnal.gov`

The 2014 P5 report [1] indicated the accelerator-based neutrino and rare decay physics research as a centrepiece of the US domestic HEP program. Operation, upgrade and development of the accelerators for the near-term and longer-term particle physics program at the Intensity Frontier face formidable challenges. Here we discuss key elements of the accelerator physics and technology R&D program toward future multi-MW proton accelerators.

---

1. Introduction

The 2014, the Particle Physics Project Prioritization Panel (P5) report [1] identified the top priority of the domestic intensity frontier high-energy physics for the next 20-30 years to be a high energy neutrino program to determine the mass hierarchy and measure CP violation, based on the Fermilab accelerator complex which needs to be upgraded for increased proton intensity. To this end, a new beam line - the Long Baseline Neutrino Facility (LBNF) – and new experiment - the Deep Underground Neutrino Experiment (DUNE), located in the Sanford Underground Research Facility (SURF) - are being planned [2]. This will be a truly international collaboration, including contributions from 150 institutions in 27 countries. The P5 physics goals require about 900 kt·MW·years of exposure (product of the neutrino detector mass, average proton beam power on the neutrino target and data taking period) and that can be achieved assuming a 40 kton Liquid Argon detector and accelerator operation with the eventual multi MW beam power. Construction of the PIP-II SRF 800 MeV linac [3] is expected to address the near-term challenges. PIP-II will increase the Booster per pulse intensity by 50% and allow delivery 1.2 MW of the 120 GeV beam power from the Fermilab's Main Injector, with power approaching 1 MW at energies as low as 60 GeV, at the start of DUNE/LBNF operations ca 2023. Extensive accelerator R&D program towards multi-MW beams has been started in the US and it has three components: demonstration of novel techniques for high-current beam accelerators at IOTA, cost-effective SC RF and high-power targetry (HPT).

2. Experimental R&D with high brightness beam at IOTA ring

Progress of the Intensity Frontier accelerator based HEP is hindered by fundamental beam physics phenomena such as space-charge effects, beam halo formation, particle losses, transverse and longitudinal instabilities, beam loading, inefficiencies of beam injection and extraction, etc. The Integrable Optics Test Accelerator (IOTA) facility at Fermilab [4] is being built as a unique test-bed for transformational R&D towards the next generation high-intensity proton facilities – see Fig. 2. The experimental accelerator R&D at the 40 m circumference IOTA ring  - see Fig.1 - with high brightness 70 MeV/c protons and 150 MeV/c electrons, augmented with corresponding modeling and design efforts has been started at Fermilab in collaboration more than two dozen universities, National and international partners [5].

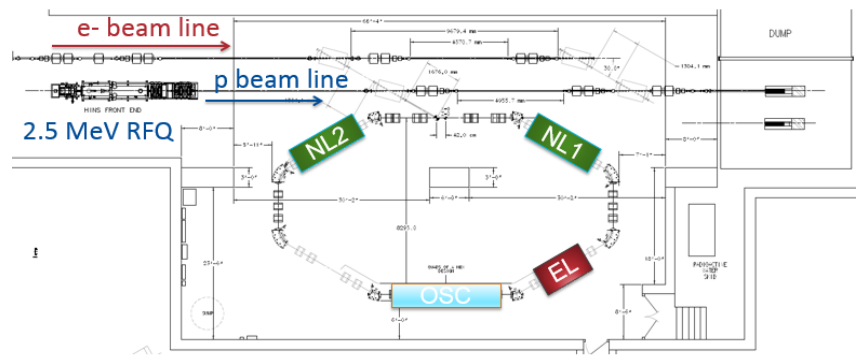

Figure 1: IOTA ring, its electron and proton injection lines and experimental areas.

The goal of the IOTA research program is to carry out experimental studies of transformative techniques to control proton beam instabilities and losses, such as *integrable optics* [6] with non-linear magnets and with electron lenses, and *space-charge compensation* with electron lenses and electron columns [7, 8] at beam intensities and brightness 3-4 times the current operational limits, i.e., at the space-charge parameter $\Delta Q_{SC}$ approaching or even exceeding 1. Several experiments are planned at IOTA: i) **Test of Integrable Optics (IO) with Electrons** with a goal to create IO accelerator lattice with several additional integrals of motion (angular momentum and McMillan-type integrals, quadratic in momentum); ii) **IO with Non-linear Magnets, Test with Protons** will demonstrate nonlinear integrable optics with protons with a large betatron frequency spread $\Delta Q_{SC}>1$ and stable particle motion in a realistic accelerator design; iii) **IO with e-lens(es), Tests with Protons** to demonstrate IO with non-Laplacian electron lenses with the electron charge distribution as $n(r)=1/(1+r^2)^2$ to obtain a large betatron frequency spread $\Delta Q_{SC}>1$ and stable particle motion in a realistic accelerator design; iv) **Space-Charge Compensation (SCC) with e-lens(es), Test with Protons** has the main goal of demonstrating SCC with Gaussian ELs with protons with a large betatron frequency spread $\Delta Q>0.5$ and stable particle motion in a realistic accelerator design. Similar *SCC tests* are envisioned *with electron columns* [8].

In 2016, the IOTA team has commissioned 50 MeV SRF electron pre-injector [5]. Operation of the IOTA ring with 150 MeV electrons in planned for 2017, and with protons in 2019.

## 3. Cost effective SRF technology

Superconducting RF is the state-of-the-art technology with an unmatched capability to provide up to 100% beam duty factor and large apertures to preserve the beam quality. In the past, the SRF R&D program has been focused on improving the accelerating gradients in "traditional" Nb structures, extending from 3 MV/m to some 35 MV/m. The demands of the Intensity Frontier accelerators shift of focus towards decreasing the costs of SRF construction and operation through: a) nitrogen doping for ultra-high Q cavities, which opens up more than a factor of two in the quality factors (Q) of bulk niobium cavities and, therefore, postential for savings in cryogenics capital and operational costs [9]; b) development $Nb_3Sn$ cavities for 4.2K operation, following the proof-of-principle demo that Nb3Sn cavities could provide the same quality factors at >4.2K as bulk niobium cavities do at 2K [10]; c) using Nb/Cu composite material and monolithic techniques of cavity manufacturing; these avenues promise a factor of >2 reduction in cavity material and manufacturing costs with performance comparable to bulk Nb cavities. It was recently shown that 1.3 GHz Nb-Cu composite based spun cavities can sustain high accelerating gradients. Fermilab, in collaboration with Cornell University, will use the existing Nb-Cu sheets to spin the cavities at INFN(Italy) or US industry (e.g. AES) to complete the 650 MHz cavities with flanges as the first step followed by scaling to 325 MHz if successful. Recent breakthrough in Nb film deposition technology allows films of unprecedented quality with the residual resistivity ratio (RRR) approaching or exceeding 200-300, which is currently the standard for bulk SRF cavities. The RF properties of these films will be tested, and after confirmation of low surface resistance of the samples, a prototype 650 MHz Nb/Cu cavity will be built and studied.

## 4. High power targetry R&D

Mega-watt class target facilities present many technical challenges, including: radiation damage, rapid heat removal, high thermal shock response, highly non-linear thermo-mechanical simulation, radiation protection, and remote handling [11]. The major goal of the envisioned R&D program for the next decade is to enable well-justified design simulations of high intensity beam/matter interactions using realistic, irradiated material properties for the purposes of designing and predicting lifetimes of multi-MW neutrino and muon target components and systems.

This requires: a) irradiated material properties to be measured/evaluated for relevant targetry materials over a range of temperatures (300 – 1300 K), radiation damage (0.1 – 20 DPA (Displacements Per Atom)) and relevant helium production rates (500 – 5000 atomic parts per million/DPA); b) thermal shock response to be evaluated for relevant targetry materials over a range of strain rates (100 – 10000 $s^{-1}$); c) development and validation of simulation techniques to model material response to beam over the time of exposure (accounting for accumulation of radiation damage and high spatial gradients); d) development of enabling technologies in target materials, manufacturing techniques, cooling technologies, instrumentation, radiation protection, and related systems to meet the targetry challenges of multi-MW and/or high intensity (> 500 MW/$m^3$ peak energy deposition) requirements of future target facilities.

*Radiation damage studies* include investigations of materials of high interest (currently graphite, beryllium, tungsten and titanium alloys) under the RaDIATE R&D program [12]. The most major of these activities involve Post-Irradiation Examination (PIE) of previously irradiated materials recovered from spent target components (e.g. NuMI proton beam window), low-energy ion and high energy proton irradiations at available beam facilities (e.g. Brookhaven Linac Isotope Producer [13]), and experiments designed to help correlate low energy ion irradiations to high energy proton irradiations.

*Thermal shock response studies* include in-beam thermal shock experiments of various grades of commercially pure beryllium at the HiRadMat Facility [14] at CERN (e.g. HRMT-24, "BeGrid" [15]) and high strain rate testing of candidate materials to develop strength and damage models.